Searching for refractory plasmonic materials: the structural and optical properties of Au$_3$Zr intermetallic thin films

*Hugh Littlehailes, William R Hendren, Stacey Drakeley, Robert M Bowman, and Fumin Huang\**

*Queen's University Belfast, School of Mathematics and Physics, University Road,*

*Belfast, BT7 1NN, United Kingdom*

*Corresponding Author: f.huang@qub.ac.uk

Abstract

Optical properties of refractory intermetallic thin films of Au$_3$Zr were experimentally investigated for the first time, which show distinctive plasmonic properties in the visible and near infrared region. The films were fabricated through DC magnetron sputtering at various deposition temperature ranging from room temperature to 427°C and annealed at different vacuum levels. Both the structural and optical properties are found to be critically dependent on deposition temperature and anneal conditions. Films deposited between 205-320°C are shown to exhibit lower negative permittivity and better thermal stability, which could be linked to specific crystalline orientations. The films are stable when annealed at $10^{-8}$ Torr, but are partially oxidized when annealed at $10^{-6}$ Torr, suggesting oxidization could be a restricting issue for high-temperature applications in ambient environment.

I. INTRODUCTION

Originating from the resonant oscillation of electrons in metallic materials when stimulated by incident electromagnetic radiation, surface plasmons allow for the transmission of energy of the incident light below the diffraction limit, when coupled at the metal-dielectric interface [1]. In systems of coupled nanoparticles, optical energy can be concentrated within extremely small cavities in the order of nm$^3$ [2-4], billions of times more compact in volume than the focus spot of a conventional optical microscope. These extraordinary capabilities have found applications in a breadth of fields including subwavelength imaging [5], energy harvesting [6], cancer therapy [7], information technology [8], and data storage [9]. Noble metals are the most popular plasmonic materials, especially gold and silver, due to their relatively low loss in the visible and near infrared region. However, the softness of noble metals and the reduction of melting points at the nanoscale make them vulnerable for applications in high-temperature environment, such as in heat-assisted magnetic recording (HAMR), where the working temperature is above 400°C [9]. This causes Au-based near-field transducer to deform, which is one of the major obstacles establishing the reliability of HAMR technology [10]. Apart from HAMR, many other high-temperature technologies and applications also have strong demand of the development of refractory plasmonic materials, such as thermoelectrics [11], thermophotovoltaics [12], water splitting [13] and nonlinear optics [14]. This has prompted great interest in the search for refractory plasmonic materials that can work in harsh environment of elevated temperature [15, 16].

Previous work has looked at various options from metallic silicides, nitrides, conducting oxides, and intermetallic alloys [17-19]. Among these, intermetallic alloys represent an interesting new paradigm of refractory plasmonic material. It is predicted that many intermetallic alloys could exhibit promising plasmonic properties [20, 21]. However, so far very little experimental work has been demonstrated. In this investigation we alloyed Au with Zr, one of the refractory metals of high melting temperature, which is also highly resistant to corrosion. We focus on the $Au_3Zr$ binary alloy, which has a melting temperature up to 1610°C [22]. Keast *et al.* performed density-of-states calculations of the imaginary part of permittivity of Au-Zr compound (87.5at%Au-12.5at%Zr), which shows a strong peak around 2.4 eV just below the interband transition of Au [20]. However, to date the optical properties of Au-Zr intermetallic alloys have not been established. The results presented here provide the first detailed experimental study of the optical properties of Au-Zr intermetallic compound.

## II. MATERIALS AND METHODS

Thin films of $Au_3Zr$, close to 100 nm thick, were co-sputtered at an argon process pressure of 0.8 mTorr using DC magnetrons from elemental gold and zirconium targets of purity better than 99.99% and 99.5% respectively in a cryo-pumped UHV deposition chamber with base pressure below $10^{-8}$ Torr. A set of films were fabricated at different substrate deposition temperature ranging from room temperature (RT) to 427°C. The thin films were deposited onto pre-cleaned Si wafers, with 300 nm oxide layer. There were two exceptions. Films deposited at 320°C and 415°C were fabricated on transparent $SiO_2$ wafers. This was intended to identify any changes on the undersides of these films during annealing processes. No adhesion layers were applied between the substrate and the deposited film.

The samples were left in the process chamber to anneal in-situ at 450°C for 1 hour, immediately after deposition, to relieve thermal stress and promote grain growth. Comparative samples, deposited at RT and 439°C were also fabricated, without the in-situ anneal. Structural properties were characterised by XRR and XRD on a Bruker D8 Discover X-ray diffractometer (Cu K$_\alpha$, $\lambda$ = 1.5418 Å) using Bruker Leptos and EVA software for data analysis. Au$_3$Zr forms an orthorhombic structure, belonging to the Cu$_3$Ti prototype [23] and the space group of Pmmn (59) with lattice parameters of a = 6.062 Å, b = 4.865 Å, c = 4.785 Å [20]. The major peaks in the samples were identified by peak fitting using these lattice parameters and space group.

The resistivity of the films was determined by using a Jandel 4-point probe. The optical properties of the structures were determined by the use of a J A Woollam Co. Inc, EC-400 spectroscopic ellipsometer, with an M-2000VI light source, operating with a spectral range from 370 nm to 1690 nm. The reflection spectra were measured using an Olympus BX51 optical microscope with a 5× objective (NA=0.1). The theoretical reflection spectra of the films were calculated from Fresnel's equation of reflectance at normal incidence, which is a close approximation of the experimental condition. To establish the thermal stability the films were annealed for a second time in a vacuum oven that had been purged with argon gas. The temperature was raised to 497°C at a rate of 3°C/min, and dwelled at this peak for 1 hour, at a chamber pressure of $1.5\times10^{-6}$ Torr.

III. RESULTS AND DISCUSSIONS

3.1 Structural properties

The thickness of the films was between 90-95 nm. Fig.1 shows the XRD spectra (the intensity is normalised to the maximal peak between 32° and 42°) of the films.

From Fig.1 it can be seen that the films are preferentially aligned along one primary crystalline orientation, with a small proportion of other textures, a trend to be expected given the lack of lattice matching between the deposition surface and the deposited film, which has been suggested for polycrystalline samples to result in higher resistivities as a result of grain-boundary effects [24, 25]. The most prominent peaks for the majority of films are the Au$_3$Zr (002) and Au$_3$Zr (121) orientations, with the lower temperature films exhibiting greater preference for the former, and the higher temperature films for the latter. The transition from the dominance of Au$_3$Zr (002) orientation to the Au$_3$Zr (121) orientation was observed in the films of 205 and 320 (for simplicity, hereafter we will simply represent the films as the number of deposition temperature, omitting the °C unit), identifiable by the greater relative proportion of the latter orientation compared to the films deposited below 205°C.

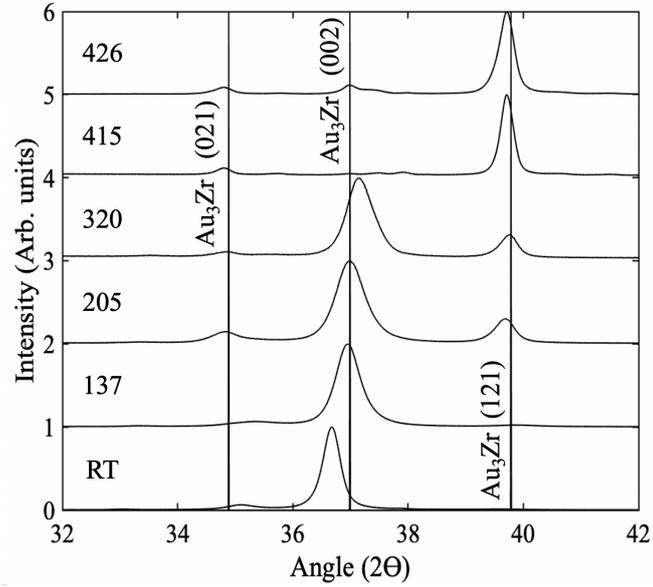

**Fig.1** XRD spectra of Au$_3$Zr films fabricated at different deposition temperatures denoted by the numbers (in the unit of Celsius). Solid lines indicate the expected peak positions calculated from the lattice parameters in literature

Figure 1 shows that both peak position and broadness change with deposition temperature, which is caused by combined effects of strain and grain size. The effects can be assessed by applying the Williamson-Hall analysis on the XRD spectra [26, 27],

$$\beta_{hkl} \cdot \cos\theta = 4\varepsilon \cdot \sin\theta + \frac{K\lambda}{D} \qquad (1)$$

$\beta$ is the width of the diffraction peak in radian, $\varepsilon$ is the strain, $D$ is the average size of grain, $\lambda$ is the wavelength of X-ray, $\theta$ is the Bragg diffraction angle, and $K$ is the Scherrer constant in the order of unity.

Though the XRD spectra are dominated by one primary peak, a weak secondary peak is identifiable in each spectrum. We chose the two most significant peaks from each diffraction spectrum in Fig.1. The results (see Supplementary Information Fig.S1 for details) are plotted as $\beta\cos\theta$ vs $4\sin\theta$ in Fig.2a.

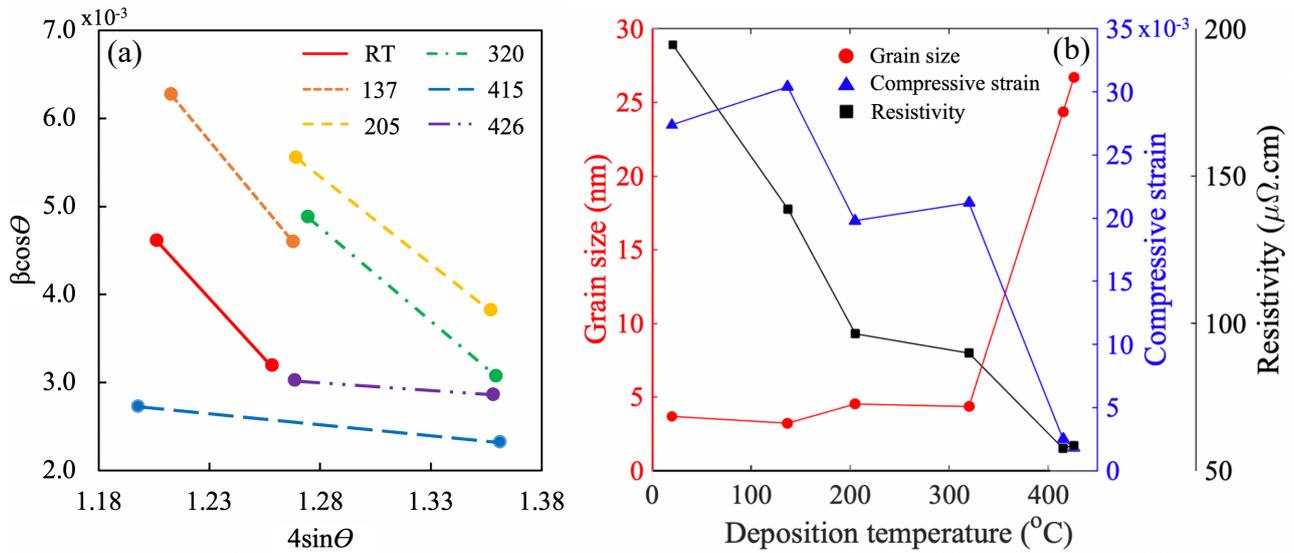

**Fig.2** Williamson-Hall analysis of the XRD results. (a) Williamson-Hall plots of films deposited at various temperatures denoted by the numbers (°C). (b) The average grain size (circles), strain (triangles) and resistivity (squares) of the $Au_3Zr$ films as a function of deposition temperature

The results of grain size (assuming Scherrer constant $K$ =0.9) and strain (the gradients of lines in Fig.2a are negative, indicating compressive strains) are displayed in Fig.2b. The strain decreases and the grain size increases with deposition temperature, which are consistent with general trends observed in thermally deposited thin films [28]. Films deposited at temperatures below 320°C have an average grain size of about 4 nm. However, the 415 and 427 films have significantly larger grain sizes of 24.4 nm and 26.7 nm, respectively. It could be possible that different crystalline orientations lead to the growth of different sizes of grains. For these two films, the dominant orientation is $Au_3Zr$ (121), while those deposited at lower temperatures are predominantly $Au_3Zr$ (002) orientation. We further measured the resistivity of the films, which are shown in Fig.2b (squares). It is seen that the resistivity decreases with deposition temperature, consistent with the trend of increasing grain size. Larger grains reduce the scattering loss of electrons at grain boundaries, thus have smaller resistivity.

3.2 Optical properties

We characterized the optical properties of the thin films through spectroscopic ellipsometry. The results are displayed in Fig.3, including both the real and imaginary components of the refractive index (Fig.3a, b) and permittivity (Fig.3c, d). The optical properties of the films can be approximately segregated into three groups: RT and 137, 205 and 320, 415 and 427. Each pair has similar optical properties, which is largely in correspondence with the textures revealed by the XRD spectra shown in Fig.1. This suggests that the optical properties of thin films are significantly impacted by crystal orientations. Most films exhibit negative real part permittivity in the visible range between 450-800 nm. In particular, the films of 205 and 320 show substantial negative real part permittivity across the entire visible and near infrared region, which are comparable to that of titanium nitride [17], one of the most promising refractory plasmonic materials. The imaginary parts of the permittivity increase almost linearly with wavelength, which are also impacted by deposition temperature, but are not as dramatic as the real parts.

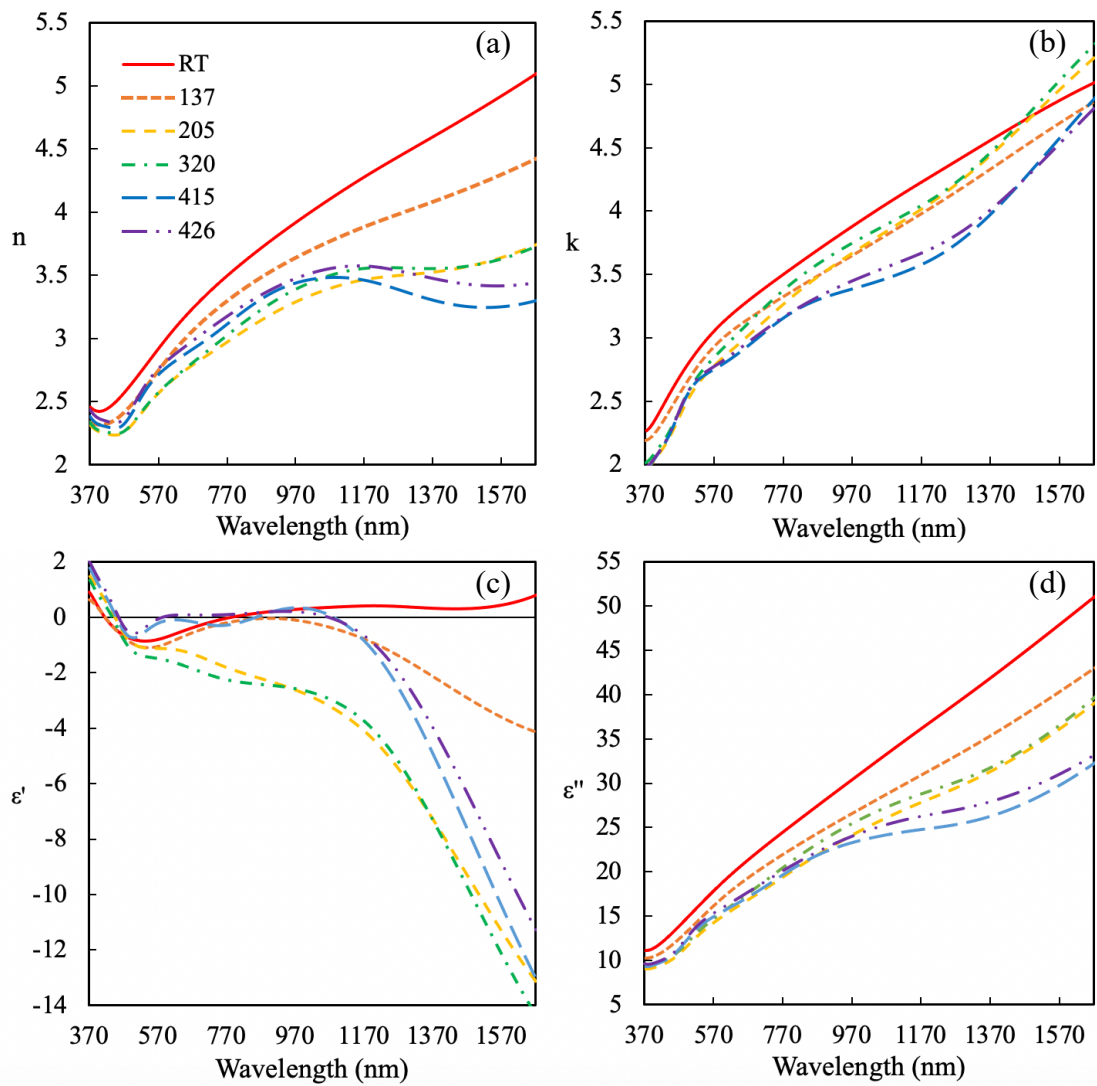

**Fig.3** Optical properties of $Au_3Zr$ thin films deposited at various temperatures. (a-b) Real (n) and imaginary (k) part of refractive index. (c-d) Real ($\varepsilon'$) and imaginary ($\varepsilon''$) part of permittivity

To further characterize the optical properties of the thin films, we measured the reflectance spectra of the films using a 5× low magnification objective and compare the spectra with theoretically calculated ones based on the Fresnel equations of normal incidence, using the refractive indices measured from the ellipsometry. The results are shown in Fig. 4. There is good agreement between experimental and theoretical curves. The maximum reflectance of these films is between 60-70% (@1000 nm), lower than that of Au which is above 90% in the near infrared.

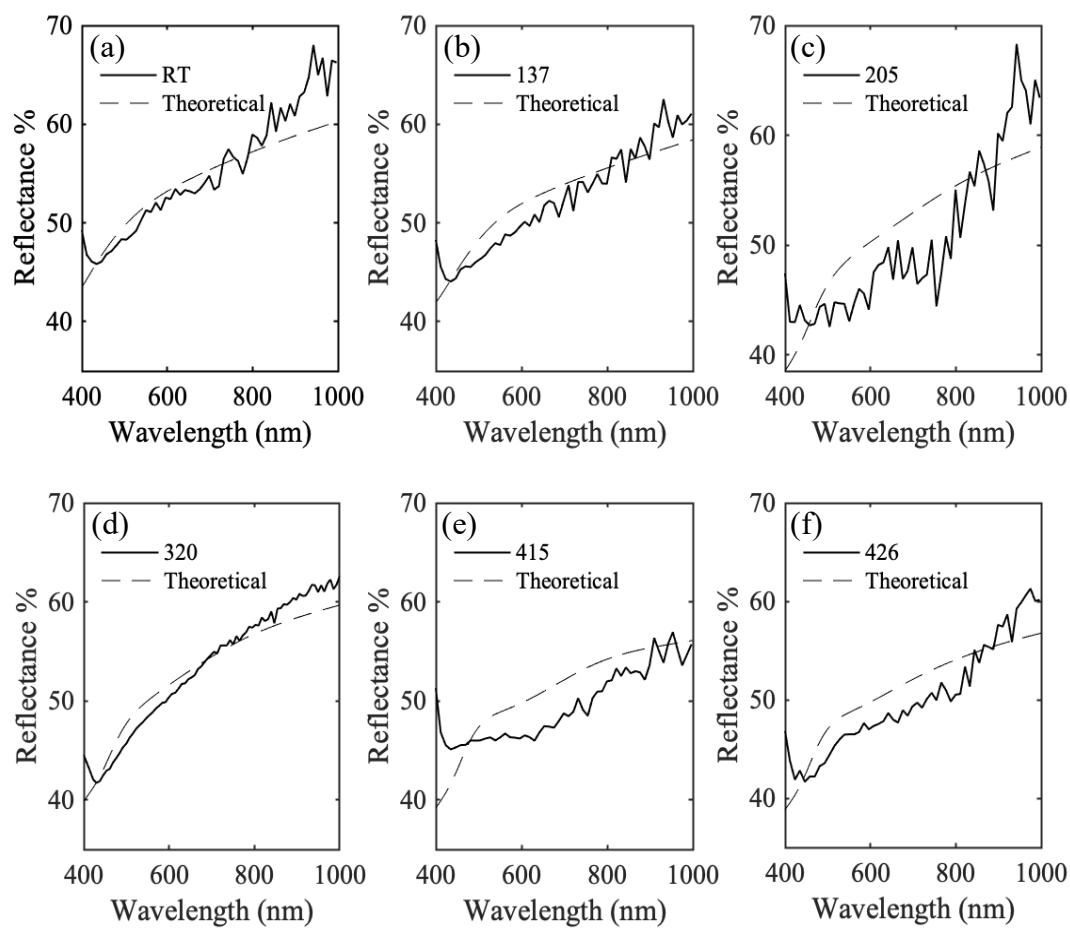

**Fig.4** Measured (solid) and calculated (dashed) optical reflectance spectra of $Au_3Zr$ thin films fabricated at various deposition temperatures

3.3 Annealing effects

We went on to investigate how anneal affects the films' structural and optical properties. Fig.5a compares RT films with and without the in-situ anneal. Very little change was noted in the XRD spectra. Fig.5b compares the XRD results of 426 films with and without the in-situ anneal (the deposition temperature of the sample without in-situ anneal was slightly higher at 439°C). The XRD spectra remain similar with and without in-situ anneal. However, more pronounced changes were noted on the two small peaks Au$_3$Zr (021) and Au$_3$Zr (002), which became weaker after the in-situ anneal with regard to the dominant Au$_3$Zr (121) peak.

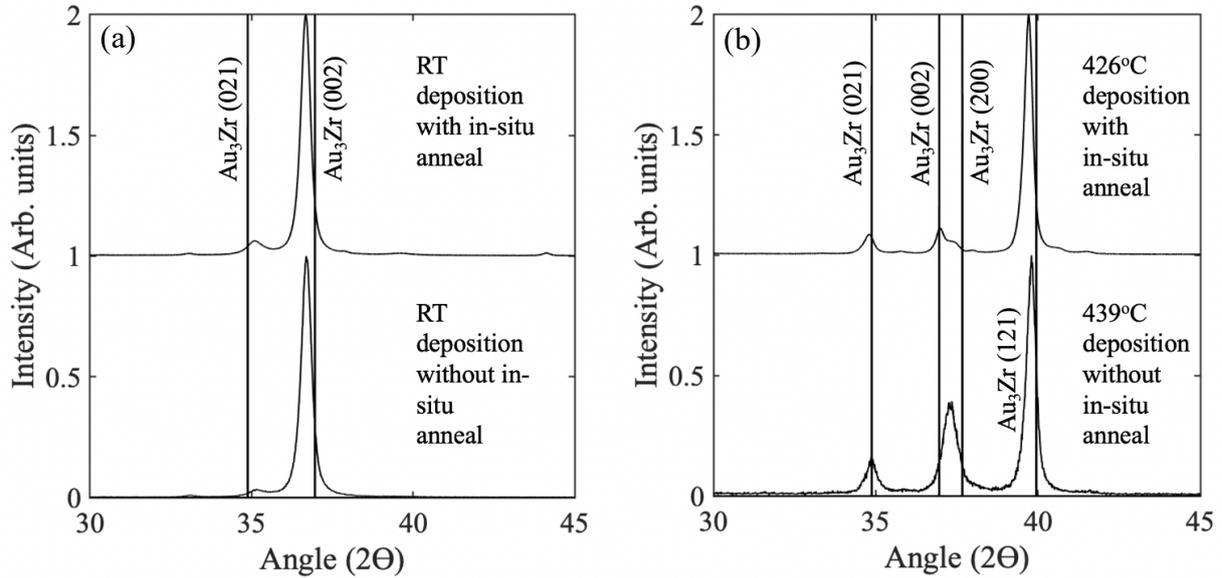

**Fig.5** Effects of in-situ anneal. Comparison of XRD spectra of Au$_3$Zr films deposited at (a) room temperature (RT) and (b) 426°C and 439°C, with and without in-situ anneal, respectively

Fig.6 compares the refractive indices for the same films as in Fig.5, fabricated with and without in-situ anneal, to investigate the impacts of in-situ anneal on the optical properties of the films. Two samples deposited at RT (Fig.6a) and the other two deposited at a higher temperature of 426°C and 439°C(Fig.6b). We can see that in-situ anneal has minimal impacts on the optical properties, which are almost identical to those without the anneal. The small difference in optical properties with in-situ anneal possibly reflects a coalesce of small grains into larger ones.

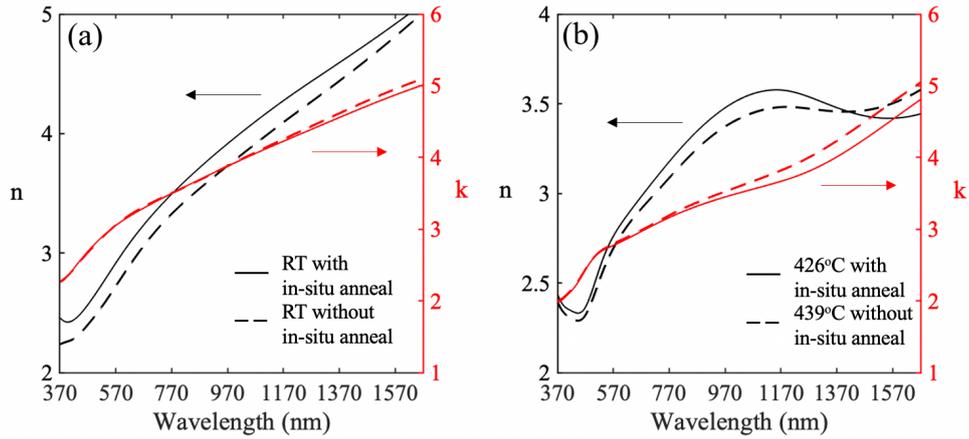

**Fig.6** The effects of in-situ anneal on the optical properties. Real and imaginary refractive index of Au$_3$Zr films with and with-out the in-situ anneal. (a) films deposited at RT, (b) films deposited at 426°C (in-situ annealed sample) and 439°C (without in-situ anneal)

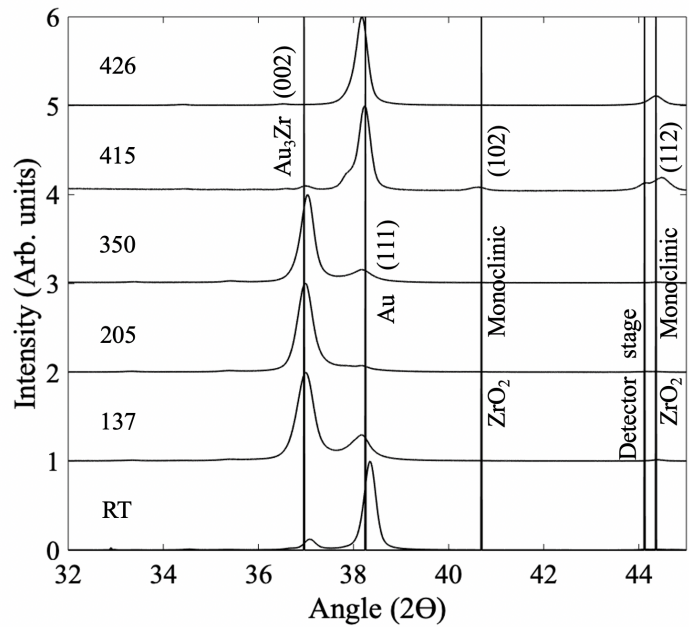

**Fig.7** XRD spectra of Au$_3$Zr films annealed in a vacuum oven (1.5×10$^{-6}$ Torr) at 497°C for 1 hour. The number of each spectrum indicates the deposition temperature (°C)

To further test the stability of the films, we annealed the films in an oven of a lower vacuum level of 1.5×10$^{-6}$ Torr at 497°C for one hour, after the films had been annealed in-situ within the sputtering chamber. The XRD spectra of the films after the second anneal are shown in Fig.7. One striking change is that the Au$_3$Zr (121) peak which is dominant in the films of 415 and 427 (Fig.1) has been completely eradicated. A new dominant Au

(111) peak appears (Au clusters are clearly visible on sample surface under optical microscope, see Supplementary Information Fig.S2), accompanied by the appearance of a weak $ZrO_2$ (112) peak. This suggests $Au_3Zr$ films were partially oxidized with element/phase segregation by the trace oxygen in the oven. Interestingly, films deposited at different temperatures show different levels of oxidation resistance. As shown in Fig.7, films of RT and 415 and 426 are easily oxidized, showing strong Au (111) peaks, while the films of 137, 205 and 320 are more resistant, only showing minute Au (111) peaks with barely visible $ZrO_2$ peaks. As shown in Fig.1, films of 415 and 426 are primarily dominated by $Au_3Zr$ (121) peaks and films of 137, 205 and 320 are primarily dominated by $Au_3Zr$ (002) peaks. This suggests that different crystal orientations could have different oxidation resistance. Many other refractory plasmonic materials, such as TiN and ZrN [29, 30], are also susceptible to the issue of oxidation. The results imply that exploring samples of different crystalline orientations could be a way worth trying to improve the resilience of oxidation.

## IV. CONCLUSION

Here we for the first time fabricated and established the optical properties of $Au_3Zr$ intermetallic thin films. We demonstrate that $Au_3Zr$ intermetallic thin films can be fabricated by magnetron co-sputtering of elemental Au and Zr at a ratio of 3:1. The films clearly show plasmonic properties in the visible and near infrared region, but both the structural and optical properties are critically dependent on deposition temperature. Films deposited between 205 - 320°C are found to produce best plasmonic properties as well as the most resistance to oxidation, which could be linked to specific crystalline orientations. The films are thermally stable when annealed in an ultrahigh vacuum chamber with a base pressure of $10^{-8}$ Torr, but are partially oxidized when annealed at $1.5\times10^{-6}$ Torr. Intermetallic alloys contain a large group of materials with high melting temperatures, which could be promising candidates of novel refractory plasmonic materials, however, till date the optical properties of intermetallics remain largely unexplored. Here we provide the first attempt to experimentally establish the plasmonic properties of $Au_3Zr$ intermetallic alloy, which lays a foundation for further investigations on other Au-Zr alloys and refractory intermetallics.

## DECLARATIONS

FUNDING

This work was supported by the Engineering and Physical Sciences Research Council (Grant number EP/L015323/1).

CONFLICTS OF INTEREST

The authors declare no conflicts of interest.

AVAILABILITY OF DATA AND MATERIAL

Data available on request from the authors.

CODE AVAILABILITY

Not applicable

AUTHOR'S CONTRIBUTIONS

HL performed sample fabrication and characterisations with the assistance of WRH. HL and FH prepared the manuscript. All authors contributed to the discussion of the results.

ETHICS APPROVAL

Not applicable

CONSENT TO PARTICIPATE

Not applicable

CONSENT FOR PUBLICATION

Not applicable

# Supplementary Information

Searching for refractory plasmonic materials: the structural and optical properties of $Au_3Zr$ intermetallic thin films

*Hugh Littlehailes, William R Hendren, Stacey Drakeley, Robert M Bowman, and Fumin Huang**

*Queen's University Belfast, School of Mathematics and Physics, University Road,*

*Belfast, BT7 1NN*

*Corresponding Author: f.huang@qub.ac.uk

| Film | Most likely structure | Peak position (degrees) | Width (degrees) |
|---|---|---|---|
| Room temperature | Au$_3$Zr (021) | 35.11 | 0.360 |
|  | Au$_3$Zr (002) | 36.67 | 0.384 |
| 137°C | Au$_3$Zr (021) | 35.30 | 0.754 |
|  | Au$_3$Zr (002) | 36.96 | 0.580 |
| 205°C | Au$_3$Zr (021) | 34.81 | 0.578 |
|  | Au$_3$Zr (002) | 37 | 0.671 |
|  | Au$_3$Zr (121) | 39.67 | 0.464 |
| 320°C | Au$_3$Zr (021) | 34.86 | 0.565 |
|  | Au$_3$Zr (002) | 37.17 | 0.589 |
|  | Au$_3$Zr (121) | 39.74 | 0.374 |
| 415°C | Au$_3$Zr (021) | 34.86 | 0.327 |
|  | Au$_3$Zr (121) | 39.80 | 0.283 |
| 426°C | Au$_3$Zr (021) | 34.78 | 0.417 |
|  | Au$_3$Zr (002) | 36.98 | 0.365 |
|  | Au$_3$Zr (121) | 39.71 | 0.349 |

**Fig.S1** The peak positions and widths of the XRD peaks of Au$_3$Zr thin films deposited at various temperatures and annealed in-situ at 450 °C for 1 hour.

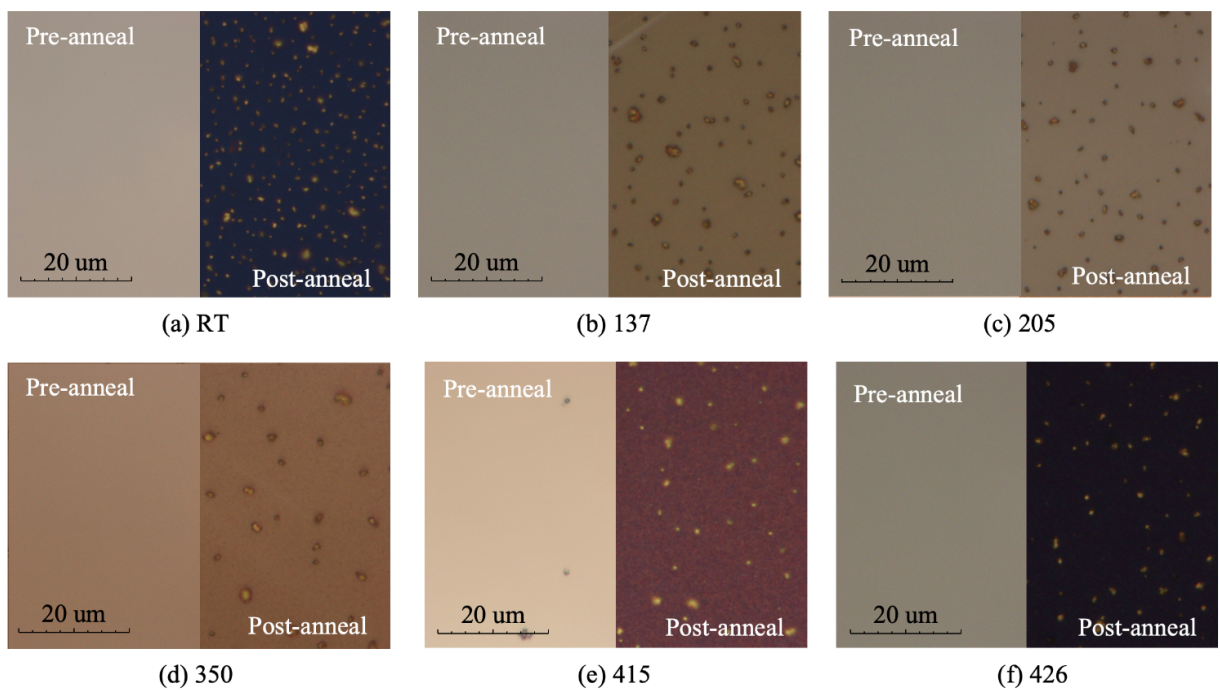

**Fig.S2** Comparison of the optical images of $Au_3Zr$ thin films before (Pre-anneal) and after (Post-anneal) being annealed in a vacuum oven of $1.5\times 10^{-6}$ Torr at 497 °C for 1 hour. The left (pre-anneal) of each image shows the films just after deposition which were annealed in-situ in an ultrahigh-vacuum chamber of $1\times 10^{-8}$ Torr at 450 °C for 1 hour. The temperature indicated is the deposition temperature of each film in °C. After the anneal inside the oven, $Au_3Zr$ films were partly oxidized, decomposing into $ZrO_2$ and Au which precipitate out and forms clusters that are clearly visible on the sample surface.